\documentstyle[12pt,aaspp4,psfig,flushrt]{article} 

\begin{document}
\bibliographystyle{apj_noskip}

\title{HIGH-SPEED OPTICAL PHOTOMETRY OF THE ULTRACOMPACT
       X-RAY BINARY 4U 1626--67}

\author{Deepto~Chakrabarty\altaffilmark{1,2}}
\affil{\footnotesize Center for Space Research, Massachusetts
   Institute of Technology, Cambridge, MA 02139}
\affil{\footnotesize deepto@space.mit.edu}

\altaffiltext{1}{NASA Compton GRO Postdoctoral Fellow}
\altaffiltext{2}{Visiting astronomer, Cerro Tololo Inter-American
  Observatory, National Optical Astronomy Observatories, operated by
  the Association of Universities for Research in Astronomy under
  contract with the National Science Foundation.}

\bigskip
\centerline{Submitted June 1, 1997 to {\sc The Astrophysical Journal}}

\begin{abstract}
Rapid {\em UBVRI\ } photometry of the ultracompact low-mass X-ray
binary (LMXB) pulsar 4U 1626--67/KZ TrA has detected 130.4 mHz (7.67
s) optical pulsations in all five bands.  The optical pulsations,
which are at the same frequency as the X-ray pulsations caused by
rotation of the highly-magnetized accreting neutron star primary, are
understood as reprocessing of the pulsed X-ray emission in the
accretion disk or on the surface of the secondary.  The optical pulsed
fraction is roughly 6\%, independent of wavelength, indicating that
the optical emission is dominated by X-ray reprocessing.  

A weaker (1.5\%) sideband, downshifted 0.395(15) mHz from the main
optical pulsation, is also present.  This is consistent with a
previously reported sideband which was downshifted 0.4011(21) mHz from
the main pulsation, corroborating the 42-min binary period proposed by
Middleditch et al. (1981, ApJ, 244, 1001).  A 0.048 Hz optical
quasi-periodic oscillation (QPO), corresponding to a previously
reported X-ray feature, was also detected in some of the observations,
with a fractional RMS amplitude of 3--5\%.  This is the first
measurement of an optical QPO in an X-ray binary pulsar.

I discuss constraints on the nature of the mass donor and show that
mass transfer via a radiatively-driven wind is inconsistent with the
optical data.  I also review the basic theory of X-ray-heated
accretion disks and show that such models provide a good fit to the
optical photometry.  If the effective X-ray albedo of LMXB accretion
disks is as high as recently reported ($\eta_d\gtrsim 0.9$), then the
optical data imply a distance of $\sim$8 kpc and an X-ray luminosity
of $\approx 10^{37}$~erg~s$^{-1}$.

\end{abstract}

\keywords{accretion, accretion disks --- binaries: close --- 
pulsars: individual: 4U 1626--67 --- stars: low-mass ---
stars: neutron}

\section{INTRODUCTION}

The optical emission from low mass X-ray binaries (LMXBs) is generally
dominated by the reprocessing of X-rays in the accretion disk and/or
the mass donor (see van Paradijs \& McClintock 1995 for a review). The
time history of the optical emission is thus a convolution of the
X-ray intensity history with a function representing the spatial
distribution of matter in the system.  The most spectacular examples
of this phenomenon are the optical novae accompanying soft X-ray
transients, which can cause the optical counterpart to brighten by
many magnitudes.

Of greater interest are those systems whose X-ray intensity histories
are more regularly or sharply modulated.  Many LMXBs show periodic
optical variability due to orbitally-modulated viewing of the X-ray
heated mass donor, providing constraints on the binary inclination and
the distribution of reprocessing material in the system (see van
Paradijs \& McClintock 1995 and references therein).  Quasi-periodic
oscillations (QPOs) at various frequencies have been observed in both
the X-ray and optical emission from the candidate black hole binary GX
339--4 (Motch, Ilovaisky, \& Chevalier 1982; Motch et al. 1983;
Imamura et al. 1990).  Correlated X-ray and optical bursts have been
observed from several systems, and the time delay and smearing of the
optical burst profiles with respect to the X-ray bursts have been used
to probe the location and distribution of matter near the neutron star
(Lewin, van Paradijs, \& Taam 1993).  Coherent optical pulsations have
been detected from the LMXB pulsars Her X-1 (Davidsen et al. 1972;
Middleditch \& Nelson 1981), 4U 1626--67 (Ilovaisky, Motch, \&
Chevalier 1978; Middleditch et al. 1981), and GX 1+4 (Jablonski et
al. 1997), and likewise provide an important probe of the binary
parameters and the distribution of reprocessing material in the
system.

The LMXB 4U 1626--67/KZ TrA consists of a 7.66 s pulsar accreting from
an extremely low-mass companion in an ultracompact binary with a
separation of $< 1$~lt-sec (Middleditch et al. 1981; Levine et
al. 1988).  The 7.66 s X-ray pulsations arise due to anisotropic
accretion of matter on the surface of a rotating, highly-magnetized
neutron star whose spin and magnetic dipole axes are misaligned.  The
spin frequency of the pulsar evolves on short ($|\nu/\dot\nu| \approx 6000$
yr) time scales due to accretion torques (Chakrabarty et al. 1997).
The optical counterpart has a strong ultraviolet excess (McClintock et
al. 1977), and optical pulsations are detected with the same frequency as
the X-ray pulsations (Ilovaisky et al. 1978).  The optical emission is
understood as primarily due to reprocessing of the incident X-ray flux
by material in the accretion disk and/or the binary companion's
surface (Chester 1979; McClintock et al. 1980). The system shows
strong, correlated X-ray/optical flares every $\sim1000$~s that are of
undetermined origin (Joss, Avni, \& Rappaport 1978; McClintock et
al. 1980; Li et al. 1980). A $\sim0.04$ Hz QPO has been detected in
the X-ray emission (Shinoda et al. 1990; Angelini et al. 1995), and
probably arises through an interaction between the pulsar's
magnetosphere and the inner edge of the accretion disk (Alpar \&
Shaham 1985; Lamb et al. 1985).

High-sensitivity timing of the optical pulsations detected an
additional weak, persistent pulsation in a lower-frequency sideband of
the ``direct'' (X-ray) pulse frequency (Middleditch et al. 1981).
This sideband was attributed to X-ray reprocessing on the surface of
the companion, so that the 0.4 mHz downshift is due to the different
effective pulsar rotation rate observed in the companion frame due to
the binary orbit\footnote{The same effect causes a sidereal day to be
4 min = (1/365) d shorter than a solar day.}.  The observed frequency
shift implies a 42-min prograde orbit.  X-ray timing measurements have
repeatedly failed to detect periodic pulse arrival time delays or
pulse frequency shifts due to a binary orbit, yielding an upper limit
of $a_x\sin i\lesssim 10$ lt-ms for the projected radius of the
neutron star orbit (Levine et al. 1988; Shinoda et al. 1990).  For a
42-min orbit, the X-ray timing limits imply a mass function $f_x(M)
\leq 10^{-6} M_\odot$, one of the smallest known for any stellar
binary.

This paper describes a program to confirm the persistent orbital
sideband reported by Middleditch et al. (1981) and also provides the
first detailed study of the optical wavelength dependence of X-ray
reprocessing in this system.  As part of this program, the first
example of an optical QPO in an X-ray pulsar system was discovered. 
A preliminary account of this work has appeared previously
(Chakrabarty 1996).

\section{OBSERVATIONS}

High speed multicolor optical photometry of 4U 1626--67/KZ TrA was
obtained on UT 1995 May 26--28 (MJD 49863--49865) with the 4-m Blanco
telescope at the Cerro Tololo Inter-American Observatory (CTIO) in La
Serena, Chile.  The observations used the Automated Single Channel
Aperture Photometer (ASCAP) and a dry-ice-cooled Varian VPM-159A
photomultiplier tube at the $f/7.5$ Ritchey-Chretien focus, along with
the {\em UBVRI} filter set described by Graham (1982).  The Varian
phototube, which has an InGaAsP photocathode, is sensitive over a wide
wavelength range, with a quantum efficiency of 15\% at 4000~\AA\ and
5\% at 9000~\AA.  All the observations were made through a 6.6~arcsec
circular aperture. The data were recorded at 1 ms resolution and
rebinned to 100 ms resolution for timing analysis. A log of the long
timing observations of 4U 1626--67 is given in Table 1.  Typical
target and sky count rates are given in Table 2. The Graham (1982)
photometric standard star E7-s was also observed at various airmasses
on each night as a calibration source.

Timing stability was maintained using a Kinemetrics Truetime 468-DC
GOES satellite synchronized clock.  This clock was synchronized using
NIST timing signals transmitted via the NOAA {\em Geostationary
Operational Environmental Satellite 7 (GOES/West)}, located in
geostationary orbit at 135$^\circ$ west longitude (Beehler \& Lombardi
1990).  The signals were corrected for the 267 ms mean path delay from
the NOAA ground station at Wallops Island, Virginia up to {\em
GOES/West} and down to the CTIO receiver (corresponding to a setting
of 57 ms for the propagation delay switch on the clock).  According to
the manufacturer specifications, the clock in this configuration
maintains absolute time accuracy within $\pm 1.5$~ms of UTC(NIST).
However, an independent check of this accuracy (e.g., a cross-check
with a Global Positioning System timing signal or optical pulse timing
of a well-established astrophysical time standard like the Crab
pulsar) was not available during these observations.

Some of the long timing observations contained a strong instrumental
signal at 60 Hz and its higher harmonics in the 1 ms data (see Table
1).  Whenever the 60 Hz signal was present, it was accompanied by a
second strong signal at 1.75 Hz and its higher harmonics, and the two
signals beat against each other. These high frequency contaminants
appeared intermittently in observations of several different targets.
However, their presence did not affect any of the science analysis for
4U 1626--67, all of which was confined to pulse frequencies below 1
Hz.

\section{RESULTS}

Absolute photometry of 4U 1626--67 on May 26 yielded the following
magnitude and colors: $V=18.68(15)$, $U-B = -1.20(17)$, $B-V =
0.02(19)$, and $V-R = 0.00(21)$.  The individual magnitudes are given
in Table 2. The $I$-band sky background observation was too short
to allow determination of a reliable source magnitude, but the
presence of 7.66-s $I$-band pulsations, in phase with those in the
bluer bands (see below), confirms that the source was detected in the
$I$-band as well.  Assuming that the $I$-band pulsed fraction is also
6\% (as in the other optical bands; see below), the source has
$I=17.8(3)$. Within the uncertainties, interstellar reddening can be
neglected for this high latitude source, since the total Galactic
reddening in the direction of 4U 1626--67 is $E(B-V)\approx 0.09$
(Burstein \& Heiles 1982).  The $UBV$ magnitudes reported here are
virtually identical to those measured on 1977 June 12 by McClintock et
al. (1977), and they agree within 2$\sigma$ with those measured on
1977 August 17 by Grindlay (1978).  However, the present $R$-band
measurement is 1.8(4) magnitudes brighter than previously reported by
Grindlay (1978).

The observed time series were transformed to the solar system
barycenter frame using the Jet Propulsion Laboratory DE-200 solar
system ephemeris (Standish et al. 1992). Power spectra of the
individual observations in all five bands showed strong 7.66 s
pulsations, with a mean barycentric pulse frequency of 130.4398(7)
mHz. This is consistent with the X-ray pulsation frequency measured at
the midpoint of the CTIO observations by the BATSE all-sky monitor on
the {\em Compton Gamma Ray Observatory} (Chakrabarty et al. 1997). To
facilitate a study of the pulse shape and phase relationship of the
optical pulsations at different wavelengths, a precise pulse phase
ephemeris was derived from a subset of the BATSE observations (1995
May 9 -- 1995 June 11) contemporaneous with the CTIO
observations. Based on these data, the barycentric X-ray pulse phase
during the CTIO observations is well fit by
\begin{equation}
\phi(t) = \phi_0 + \nu_0(t-t_0) + \frac{1}{2}\dot\nu(t-t_0)^2 ,
\end{equation}
where $\phi_0=0.936758$, $\nu_0=0.1304376013765$~Hz,
$\dot\nu = -6.32044\times 10^{-13}$ Hz s$^{-1}$, $t_0 =$ 1995
May 27.5 TDB, and pulse phase is defined relative to the intensity
minimum of the fundamental harmonic of the 20--60 keV BATSE pulse
profile.

The CTIO timing observations for each of the five filters were folded
according to the timing model in Equation (1). The resulting pulse
profiles, along with the BATSE 20--60 keV pulse profile, are shown in
Figure 1 relative to the BATSE pulse phase.  The mean optical pulse
and the BATSE hard X-ray pulse agree in phase to within 5\% (i.e., a
relative time delay of $\lesssim 0.4$ s), consistent with previous
measurements (McClintock et al. 1980).  Both the BATSE hard X-ray
pulse and the optical pulse shapes have a dominant sinusoidal
component, although the optical pulses clearly have additional
harmonic content as well.  The optical pulse shapes do not resemble
any of the X-ray pulse shapes in the 2--13 keV range, all of which
have an unusual and distinctive close double-peak structure (Levine et
al. 1988). Instead, the optical pulse shapes resemble the low-energy
(1--2 keV) and high-energy (15--60 keV) pulse shapes.  This is
consistent with previous observations by McClintock et al. (1980), who
concluded that the 1--2 keV X-rays were chiefly responsible for the
reprocessed optical emission.  The pulsed fraction, independent of
wavelength, is approximately 6\% (see Table 2), suggesting that the
optical emission is dominated by X-ray reprocessing. 

Figure~2 shows the Fourier power spectrum of a 2-hour $U$-band
observation (run 15U).  The spectrum has been rebinned into uniform
logarithmic frequency intervals, the mean noise level due to Poisson
counting statistics subtracted off, and the power level normalized
relative to the double-sided mean source power (e.g., Miyamoto et
al. 1994).  In addition to the 130 mHz pulsation, a 0.048 Hz QPO was
also detected. This feature was fit with a Lorentzian profile
(e.g., Shibazaki \& Lamb 1987),
\begin{equation}
P_{\rm QPO}(\nu) \propto \frac{(\Delta\nu/2)}{\pi}
     \frac{1}{(\Delta\nu/2)^2 + (\nu-\nu_{\rm QPO})^2} ,
\label{eq-qpo}
\end{equation}
where $\nu_{\rm QPO}$ is the centroid frequency of the QPO and
$\Delta\nu$ is its full width at half-maximum (FWHM). The fractional
root-mean-squared (RMS) amplitude of the QPO was computed by
integrating equation (\ref{eq-qpo}) over the FWHM of the feature and
taking the square root, yielding 3.4\% for this observation. The
QPO is also detected in the $B$ and $R$-band observations. The
best-fit parameters for the QPO features are included in Table 2. These
measurements provide the first detection of an optical QPO from an
X-ray pulsar.  A similar feature is observed at X-ray wavelengths
(Shinoda et al. 1990; Angelini et al. 1995; Chakrabarty et al. 1997,
in preparation).   For comparison, the Fourier power spectrum of the
2--6 keV data from a 71 ks observation of 4U 1626--67 with the
Proportional Counter Array on the {\em Rossi X-Ray Timing Explorer
(RXTE)} is shown in Figure~3.   This observation was made on 1996
February 11. The QPO is clearly visible in the X-ray data with an RMS
strength of 6\%, along with several harmonics of the coherent 7.66 s
pulsation.  (The higher harmonics are very strong due to the very
sharp 2--6 keV X-ray pulse shape of 4U 1626--67.)  The {\em RXTE} data
will be discussed in further detail elsewhere (Chakrabarty et
al. 1997, in preparation).  

To study the detailed structure of the optical timing properties 
close to the pulsation frequency, an overresolved power spectrum was
constructed by computing the Fourier power
$P_{k+\epsilon}=|H_{k+\epsilon}|^2$ at fractional bin intervals
according to 
\begin{equation}
 H_{k+\epsilon} =  {\sum_{j=0}^{N-1}  h_{j} e^{2 \pi ij(k+\epsilon)/N}}
	= {\sum_{j=0}^{N-1}  y_{j}^{(\epsilon)} e^{2 \pi ijk/N}}
\end{equation}
with $0<\epsilon<1$, where $y_{j}^{(\epsilon)}$ is just the original
time series $h_j$ multiplied by complex phase factors $e^{2 \pi ij
\epsilon /N}$. Note that while the $N$ shifted power spectrum bins
$\{P_{k+ \epsilon}\}$ are statistically independent of each other,
they are {\em not} independent of the original power spectrum
$\{P_k\}$ (although the covariances are easily computed; see, e.g.,
Jenkins \& Watts 1968).  Generalizing this process, one can construct an
$n$-times overresolved power spectrum by interleaving the bins from
the original power spectrum and $n-1$ frequency-shifted power spectra.
Operationally, this is equivalent to padding the original $N$-point
zero-mean time series with $(n-1)N$ zeros and computing the power
spectrum of the new $nN$-point padded time series. Use of an
oversampled power spectrum gives more uniform frequency sensitivity at
the expense of independent frequency bins, recovering Fourier phase
information which is normally lost when constructing the power
spectrum from the Fourier amplitudes.

Figure 4 shows a 5-times overresolved power spectrum of 4U 1626--67 in
the immediate vicinity of the main pulsation.  The power spectrum has
been normalized relative to the mean noise power.  To improve the
signal-to-noise ratio, the power spectrum shown is the average of
power spectra from three separate 1.8-hr observations (runs 15U, 17B,
and 23R). In addition to the fundamental pulsation at 130.4398(37)
mHz, several other features are present. In considering these
features, it is important to recall that the peaks due to periodic
signals in a finite-length power spectrum are always accompanied by
sidelobes. This can be understood by considering a finite time series
as the product of an infinite data stream and a rectangular (boxcar)
observation window function (equal to unity during the observation and
zero elsewhere). By the Fourier convolution theorem, the Fourier
transform of this product is equal to the convolution of the Fourier
transforms of the infinite data stream and the observation window
function (e.g., Press et al. 1992). Thus the power at frequency $\nu$
due to a sinusoidal signal with frequency $\nu_0$ in an ungapped
observation of length $T$ is
\begin{equation}
P(\nu) = P_0\left[\frac{\sin \pi(\nu_0-\nu)T}{\pi(\nu_0-\nu)T}\right]^2,
\end{equation}
where $P_0$ is the power at the signal frequency. The dotted curve in
Figure 4 shows the fit of this function to the data (with the expected
mean noise level of unity added), assuming that $\nu_0=130.4398$ mHz
corresponds to the main pulsation peak. The expected sidelobe
structure almost exactly accounts for the two symmetric sidelobes on
either side of the main peak.

However, after accounting for the expected sidelobe structure of the
main pulsation, a residual power excess is clearly evident at
$\nu=130.045(14)$ mHz, downshifted 0.395(15) mHz from the main
pulsation peak. This is consistent with the 0.4011(21) mHz shift for
the lower sideband reported by Middleditch (1981).  The amplitude of
the sideband relative to the main pulsation is 25\% (compared to 20\%
measured by Middleditch et al. 1981), corresponding to a sideband
pulsed fraction of 1.5\%.  To evaluate the statistical significance of
this excess power, note that the powers in an average of three
background-dominated power spectra obey (within a constant scaling
factor) a $\chi^2$-distribution with six degrees of freedom. For a
power spectrum normalized such that the mean noise power is unity, the
probability of a noise fluctuation in a single bin
exceeding a threshold power $P_{\rm thresh}$ is
\begin{equation}
\Pr(P>P_{\rm thresh}) = \frac{1}{16}\int_{6P_{\rm thresh}}^\infty
	t^2 e^{-t/2} dt .
\end{equation}
Thus, the observed excess power of 5.9 has a probability of $3.6\times
10^{-6}$ of being due to a random fluctuation, making it significant
at the 4.6$\sigma$ level.  Accounting for the fact that all bins in
the range 0.129--0.132 Hz were searched reduces the statistical
significance to 4.0$\sigma$. It is especially striking to note that
the frequency {\em shift} of the sideband agrees with that reported by
Middleditch et al. (1981) despite the fact that the actual value of
the main pulsation frequency (corresponding to the pulsar's spin) has
changed from 130.26 mHz to 130.44 mHz in the 15 years between the two
observations (Chakrabarty et al. 1997). Our sideband detection
strongly corroborates the 42-min binary period reported by Middleditch
et al. (1981).  Unfortunately, the signal-to-noise ratio of this detection
is insufficient to permit a more detailed investigation of the binary
parameters.

\section{DISCUSSION}

\subsection{Multiwavelength timing signatures as probes of LMXBs}

QPOs in accretion-powered pulsars are widely believed to arise from
interactions between the pulsar magnetosphere and the inner edge of an
accretion disk.  In the beat frequency model (Alpar \& Shaham 1985;
Lamb et al. 1985), the QPO arises when inhomogeneous ``clumps'' of
material at the inner edge of the disk are captured by the
magnetosphere and accreted by the pulsar, producing a broad power
spectral feature at the beat frequency $\nu_{\rm QPO}=\nu_K-\nu_{\rm
spin}$ between the Keplerian frequency at the inner disk edge and the
pulsar's spin frequency. [An alternative explanation, the Keplerian
frequency model of van der Klis et al. (1987), predicts $\nu_{\rm QPO}
= \nu_K$.  However, it can only explain QPOs where $\nu_{\rm QPO} >
\nu_{\rm spin}$, and thus is not applicable in 4U 1626--67.] Thus, 
the 0.048 Hz X-ray QPO in 4U 1626--67 bolsters the premise that this is a
disk-fed binary. 

The discovery of an optical QPO at the same frequency shows that a
significant fraction of the optical luminosity from the binary simply
mirrors the X-ray emission, and must therefore result from X-ray
reprocessing.  As noted by Middleditch et al. (1981), the existence of
two distinct, persistent, coherent optical pulsation frequencies 
suggests that the reprocessing occurs in two separate sites, most
likely the surface of the mass donor and the accretion disk.
Simultaneous X-ray and optical timing of these pulsations and the QPO
may therefore provide a powerful probe of the geometry of the binary
and physics of the accretion disk (e.g., Arons \& King 1993).  

Unfortunately, the BATSE hard X-ray measurements acquired simultaneously
with the optical observations require $\gtrsim 1$ d to detect the X-ray
pulsations, making them useless for a correlative study on short
time scales.  However, more sensitive X-ray timing missions (e.g. {\em RXTE})
are well suited to such a task if the observations can be coordinated
with ground-based optical observations.   A comparison of Figures 2
and 3 gives a flavor of the kind of data sets that can be directly
intercompared if measured simultaneously. Ultraviolet timing
observations with the {\em Hubble Space Telescope} may also provide an
important probe, as the reprocessing in the disk should be more
efficient at these wavelengths than in the optical (Arons \& King
1993; Anderson et al. 1997).   More generally, time-resolved
ultraviolet/optical/infrared photometry may provide an important
test for candidate multiwavelength counterparts in time-variable LMXBs
whose companions remain unidentified.  

\subsection{Nature of the mass donor and the mass transfer}

While X-ray binaries and cataclysmic variables with a hydrogen-rich
mass donor cannot reach orbital periods $P_{\rm orb}\lesssim 80$ min
(Paczynski \& Sienkiewicz 1981), those with a hydrogen-depleted mass
donor can evolve to extraordinarily short ($P_{\rm orb}\sim$10 min)
orbital periods while maintaining high mass transfer rates (Nelson,
Rappaport, \& Joss 1986).  Few such systems are known, including the
six AM CVn cataclysmic variables (see Warner 1995) and the X-ray
bursters 4U~1820--30 ($P_{\rm orb}=11$ min) and 4U~1916--05 ($P_{\rm
orb}=50$ min).  The recently reported 20.6-min ultraviolet modulation
in the X-ray burster 4U 1850--087 is probably due to the same
phenomenon (Homer et al. 1996).  With a 42-min binary period, 4U
1626--67 is clearly a member of this class as well (Levine et
al. 1988).

In addition to the inferred 42-min binary period, there are several
other constraints on scenarios for the mass donor in 4U 1626--67. The
stringent X-ray timing limit ($a_x\sin i<8$ lt-ms; Levine et al. 1988;
Shinoda et al. 1990) on the projected radius of the pulsar orbit
implies 
\begin{equation}
\sin i < 7.8\times 10^{-3}\ q^{-1}\,(1+q)^{2/3}\,M_{1.4}^{-1/3}\,
    P_{42}^{-2/3} , 
\label{eq-incl}
\end{equation}
where $q=M_*/M_x$ is the mass ratio, $M_*$ is the companion mass,
$M_{1.4}$ is the neutron star mass $M_x$ in units of $1.4 M_\odot$,
and $P_{42}$ is the binary period in units of 42 min.  The {\em a
priori} probability of observing a system within the inclination limit
$i<i_{\rm max}$ is given by $(1-\cos i_{\rm max})$.

A second constraint is the assumption that the mass transfer in an
ultracompact binary is driven by angular momentum losses via
gravitational radiation (e.g., Levine et al. 1988).  The resulting
mass transfer rate will be (e.g., Verbunt \& van den Heuvel 1995 and
references therein)
\begin{equation}
\dot M = 8.3\times 10^{-8} \ M_{1.4}^{8/3}\,
      q^2\, (1+q)^{-1/3}\, \left(\frac{5}{6} + \frac{n}{2} - q\right)^{-1} \
      \ M_\odot {\rm\ yr}^{-1} ,
\label{eq-mdot}
\end{equation}
where $n$ is the power-law exponent in the companion's mass-radius
relation ($R_* \propto M_*^n$), and conservative mass transfer via
Roche lobe overflow has been assumed.  Since the 
bolometric X-ray flux from 4U 1626--67 has been measured
($F_x=2.4\times 10^{-9}$ erg cm$^{-2}$ s$^{-1}$ in the 0.5--60 keV
band; Pravdo et al. 1979), this also constrains the source
distance and X-ray luminosity $L_x=GM_x\dot M/R_x = 4\pi D^2 F_x$,
where $R_x\approx 10$ km is the neutron star radius and $D$ is the
distance to the source. 

Another constraint is a lower limit on $\dot M$ from the observed spin
history of the pulsar (Chakrabarty et al. 1997).  If the pulsar is
spinning up, then the maximum possible accretion torque on the neutron
star will occur when the magnetospheric radius $r_m$ (where the
magnetic energy density is comparable to the kinetic energy density of
the accreting matter) is close to the corotation radius $r_{\rm co}$
(where the magnetic field lines move with the local disk velocity).
This limit may be written as $N \lesssim \dot M \sqrt{G M_x r_{\rm
co}}$, where $r_{\rm co}=(G M_x /4\pi^2\nu^2)^{1/3}$ for a Keplerian
disk.  However, the torque on the neutron star can be rewritten as $N=
2\pi I_x\dot\nu$ where $I_x\approx (2/5)M_xR_x^2$ is the pulsar's
moment of inertia (Ravenhall \& Pethick 1994) and $\dot\nu$ is the
spin frequency derivative -- a directly measurable quantity.  Noting
that $\dot\nu= 8.5\times 10^{-13}$ Hz s$^{-1}$ was measured during the
spin-up of 4U 1626--67, Chakrabarty et al. (1997) thus inferred that
$\dot M\gtrsim 2\times 10^{-10}\ M_\odot$ yr$^{-1}$.  

Given the measured X-ray flux, the $\dot M$ limit requires $D\gtrsim
3$ kpc.  In view of the source's high Galactic latitude ($b=
-13^\circ$), the resulting distance out of the plane is also a 
consideration when contemplating mass donors. However, this needn't be
a strong consideration, since it is plausible for neutron star LMXBs
to have high velocities due to a supernova ``kick'', leading to a
considerable distance from the plane over the long X-ray lifetime of
the source. Indeed, the high-latitude ($b=38^\circ$) LMXB pulsar Her
X-1/HZ Her is $6.6\pm 0.4$ kpc distant, over 4 kpc out of the plane
(Reynolds et al. 1997).

The possible types of Roche-lobe-filling mass donor for 4U 1626--67
have been discussed previously by Levine et al. (1988) and Verbunt,
Wijers, \& Burm (1990).  I review and update those discussions here.
There are three possible types of Roche-lobe-filling donors for a
42-min neutron star binary.  The first possibility is a $0.02 M_\odot$
helium or carbon-oxygen white dwarf.  From equation (\ref{eq-incl}),
the binary inclination for this case must be $i\lesssim 33^\circ$,
with an {\em a priori} observation probability of 16\%.  The
mass-radius relation for such a star is $R_*\propto M_*^{-1/3}$
(Shapiro \& Teukolsky 1983), so the mass transfer rate from equation
(\ref{eq-mdot}) is $3\times 10^{-11} M_\odot$ yr$^{-1}$. This
corresponds to a distance of about 1 kpc. Although the this case has
the most favorable inclination constraint of the Roche-lobe-filling
options (due to the extremely small $M_*$), the mass transfer rate
driven by gravitational radiation alone is an order of magnitude below
the minimum mass accretion rate inferred by Chakrabarty et al. (1997).

The second Roche-lobe-filling possibility is a $0.08 M_\odot$
hydrogen-depleted star which is not fully degenerate, with $i\lesssim
8^\circ$ and an {\em a priori} probability of 1\%.  Assuming $R_*
\propto M_*$, equation (\ref{eq-mdot}) gives $\dot M = 2\times
10^{-10} M_\odot$ yr$^{-1}$, corresponding to a distance of 3 kpc.
These are identical to the lower limits on $\dot M$ and $D$ inferred
from the pulsar spin-up measurements, and thus are just consistent with the
observations.   Finally, a $0.6 M_\odot$ Roche-lobe-filling
helium-burning star is also a possible donor for a 42-min neutron star
binary. However, the small allowed inclination range ($i< 1.3^\circ$)
has an {\em a priori} observation probability of only 0.03\%, and the
expected mass transfer rate ($\dot M = 3\times 10^{-8} M_\odot$
yr$^{-1}$; Savonije, de Kool, \& van den Heuvel 1986) would require an
unphysically large distance (36 kpc) to be consistent with the
observed X-ray flux.

Angelini et al. (1995) have proposed an alternative to a
Roche-lobe-filling donor. These authors detected an unusually strong
complex of neon emission lines in the X-ray spectrum of 4U~1626--67. 
Since neon is a product of helium burning, Angelini et al. (1995) 
suggest that the mass donor is a helium-burning star which underfills
its Roche lobe, transferring matter to the neutron star via a
radiatively-driven wind.  This possibility is of particular interest
given the suggestion that the current spin-down of the pulsar is due to
accretion from a retrograde disk (Nelson et al. 1997), since formation
of such a disk is much more plausible for wind accretion than for
Roche lobe overflow.  

However, the wind from a low-mass helium-burning star would not
provide a high enough mass transfer rate to be consistent with the
X-ray data, as is now shown.  A mass of $\gtrsim 0.3 M_\odot$ is
required for core helium burning (Kippenhahn \& Weigert 1990). A $0.3
M_\odot$ helium-burning donor would fill about 25\% of its Roche lobe
(Eggleton 1983) and would have $R_*=0.07 R_\odot$ and an effective
temperature $T_*=2\times 10^4$ K (Savonije et al. 1986), yielding a
blackbody spectrum consistent with or fainter than the optical
photometry for distances $\gtrsim 4$ kpc.  More generally,
helium-burning donors with $M_* \lesssim 0.6 M_\odot$ would underfill
the Roche lobe and be consistent with or fainter than the optical
measurements for somewhat larger distances, with $R_*\lesssim 0.14
R_\odot$ and $T_* \lesssim 4\times 10^4$ K.  Assuming a
radiatively-driven wind similar to that observed in other hot stars,
the wind mass loss rate with these parameters would be $-\dot M_w
\lesssim 10^{-11} M_\odot$ yr$^{-1}$ (Abbott 1982).  Especially in
light of the poor accretion efficiency expected for a wind-fed system
($< 0.01$; e.g., Frank, King, \& Raine 1992), this is several orders of
magnitude too small to be consistent with the minimum accretion rate,
$\dot M< 2\times 10^{-10} M_\odot$ yr$^{-1}$.  Similar considerations
seem to rule out any plausible donor transferring matter via a
conventional radiatively-driven wind.  On the other hand, a
self-excited wind arising from X-ray heating of the mass donor might
be a significant contributor to the mass transfer rate, whether or not
the donor fills its Roche lobe (see Tavani \& London 1993 and
references therein).

\subsection{Origin of the optical emission}

It is instructive to consider whether a stellar spectrum can explain
the optical photometry.  The data can in fact be well fit by a 
single-temperature (stellar) blackbody spectrum, 
\begin{equation}
 F_{\nu*}=\left(\frac{R_*^2}{D^2}\right)\ \frac{2\pi
 h\nu^3/c^2}{\exp(h\nu/kT_*)-1} , 
\end{equation}
where $F_{\nu*}$ is the flux density from the star per unit frequency
$\nu$, $R_*$ is the stellar radius, $D$ is the source distance,
$h=6.6\times 10^{-27}$ erg s is Planck's constant, $c$ is the velocity
of light, $k=1.38\times 10^{-16}$ erg K$^{-1}$ is Boltzmann's
constant, and $T_*$ is the effective temperature of the star.  The
best-fit parameters are $T_*=1.8\times 10^4$ K and $R_*/D=7.3\times
10^{-13}$, so that $D = 32 (R_*/R_\odot)$ kpc.  A hydrogen-burning
star with the correct temperature (a main sequence or supergiant B
star) is obviously excluded both by the absurd ($D\gtrsim 100$ kpc)
distance required and the infinitesimal {\em a priori} probability for
the necessary binary inclination.  A $0.3 M_\odot$ He-MS star with
$D\approx 2$ kpc would have the correct temperature and flux. However,
as noted above, it would severely underfill its Roche lobe and so
probably could not transfer sufficient mass to the pulsar to explain
the observed X-ray flux.  In addition, the implied binary inclination
for a 42-min orbit has an a priori probability of $<$ 0.1\%.

However, assuming a typical white dwarf mass-radius relation,
$(R_*/R_\odot) \approx 0.01 (M_*/M_\odot)^{-1/3}$ (Shapiro \& Teukolsky
1983), we find $D\approx 1$ kpc for a $0.02 M_\odot$ white dwarf. As
noted above, such a star would just fill its Roche lobe, and the
expected mass transfer rate would be consistent with the observed
X-ray flux. For such a close binary, X-ray heating would certainly
dominate the intrinsic luminosity of the degenerate dwarf. Assuming
isotropic X-ray emission, we have
\begin{eqnarray}
T_* & \approx & \left[\frac{1}{4} \frac{R_*^2}{a^2}
     \frac{L_x(1-\eta_*)}{\pi\sigma R_*^2}\right]^{1/4} 
      = \left[\frac{L_x(1-\eta_*)}{4\pi\sigma a^2}\right]^{1/4}
\nonumber \\
    & = & 2.1\times 10^4\ P_{42}^{-1/3}D_{\rm
     kpc}^{1/2}(1+q)^{-1/6}(1-\eta_*)^{1/4} {\rm\ \ K} , 
\end{eqnarray}
where $a$ is the binary separation, $\eta_*$ is the X-ray albedo of
the white dwarf, and $F_x=2.4\times 10^{-9}$ erg cm$^{-1}$ s$^{-1}$
has been used for the X-ray flux (Pravdo et al. 1979).  Thus, X-ray 
heating of a low mass WD would naively seem to provide a plausible
model for the observed optical emission. 

Unfortunately, this is an unrealistic model. The 0.04 Hz QPO detected
in the X-ray and optical data underscores the presence of an accretion
disk (e.g., van der Klis 1995) which must contribute significant
optical luminosity.  The flux from a geometrically thin, optically
thick disk can be written as
\begin{equation}
F_{\nu d} = \frac{4\pi h\nu^3 \cos i}{c^2 D^2} \int_{r_{\rm in}}^{r_{\rm
out}} \frac{r\ dr}{\exp[h\nu/kT(r)] - 1}, 
\end{equation}
where $T(r)$ is the disk's surface temperature as a function of
mid-plane radius $r$. Given the X-ray timing limits, it is clear that
$\cos i\approx 1$.  The outer disk may be assumed to cut off sharply
at the tidal radius of the neutron star ($\approx 0.9 R_{\rm Roche}$;
Frank et al. 1992), corresponding $r_{\rm out}\approx 2\times 10^{10}$
cm for 4U 1626--67.  The inner accretion disk should be disrupted by
the pulsar's magnetosphere at roughly the neutron star's corotation
radius, $r_{\rm in}\approx r_{\rm co}= 6.5\times 10^{8}$ cm for a
Keplerian disk.  (Strictly speaking, $r_{\rm co}$ is an upper limit on
$r_{\rm in}$, but the calculation is not sensitive to this quantity
since most of the disk's optical emission comes from its outer parts.)
For an X-ray heated disk, the disk may be considered as consisting of
up to three distinct regions: an innermost region powered primarily by
internal viscous dissipation, a middle region of ``shallow'' X-ray
heating, and an outer region of ``deep'' X-ray heating (e.g.,
Cunningham 1976; Arons \& King 1993).  Let us consider each of these
regions in turn.

The standard temperature profile for an unirradiated thin accretion
disk is set by internal viscous dissipation (Shakura \& Sunyaev 1973;
Frank et al. 1992)   
\begin{equation}
T_0 = \left(\frac{3GM_x\dot M}{8\pi\sigma r^3}\right)^{1/4} =
     7100\ M_{1.4}^{1/4}\, \dot M_{-10}^{1/4}\, r_{10}^{-3/4} {\rm\ \ K} , 
\end{equation}
where $\sigma=5.67\times 10^{-5}$ erg cm$^{-2}$ K$^{-4}$ s$^{-1}$ is
the Stefan-Boltzmann constant, $M_{1.4}$ is the neutron star mass in
units of 1.4 $M_\odot$, $\dot M_{-10}$ is the mass accretion rate
in units of $10^{-10}\ M_\odot$~yr$^{-1}$, and $r_{10}$ is
the mid-plane radius in units of $10^{10}$ cm.  At sufficiently small
radii, this internal heating will dominate over X-ray heating in
setting the disk temperature.  Beyond a critical radius, however,
X-ray heating of the disk surface will modify the temperature profile
according to
\begin{equation}
T_{\rm irr}=(T_0^4 + T_x^4)^{1/4} .
\end{equation}
Assuming that the disk is irradiated by a central X-ray point
source, then we can write $T_x$ as
\begin{equation}
T_x = \left[\frac{L_x(1-\eta_d)}{4\pi\sigma r^2}
   \cos\psi\right]^{1/4} \approx \left[\frac{L_x(1-\eta_d)}
   {4\pi\sigma r^2} \left(\frac{dH}{dr} - \frac{H}{r}\right)\right]^{1/4},
\label{eq-tx}
\end{equation}
where $\eta_d$ is the X-ray albedo of the disk, $\psi$ is the angle
between the normal to the disk surface and the vector from the neutron
star, and $H$ is the disk's scale height.   Thus, we see that the
temperature profile due to X-ray heating is sensitive to the
functional form of $H(r)$ through the approximation for $\cos \psi$
(which assumes $H\ll r$).

In the shallow X-ray heating region, the surface temperature profile is
set by X-ray heating but the {\em central} temperature of the disk is
still set by internal viscous dissipation.  Thus, the disk thickness
has the usual value for a standard unirradiated thin disk (Shakura \&
Sunyaev 1973; Frank et al. 1992),
\begin{equation}
H = 1.2 \times 10^8\ \alpha^{-1/10}\, \mu^{-3/8}\, M_{1.4}^{-3/8}\, 
    \dot M_{-10}^{3/20}\, r_{10}^{9/8} {\rm\ cm} ,
\label{eq-h98}
\end{equation}
where $\alpha$ is an order unity dimensionless parametrization of the
kinematic viscosity $\nu_{\rm visc} = \alpha c_s H$, $c_s$ is the
isothermal sound speed, and $\mu$ is the mean molecular weight in
units of the hydrogen atomic mass. (Note that $\mu=1$ for neutral
hydrogen, 1/2 for ionized hydrogen, 4 for neutral helium, 2 for
singly-ionized helium, and 4/3 for doubly-ionized helium.)  We
thus find 
\begin{equation}
T_x = 1.2\times 10^4\ \alpha^{-1/40}\, \mu^{-3/32}\, (1-\eta_d)^{1/4}\,  
     M_{1.4}^{5/32}\, \dot M_{-10}^{23/80}\, r_{10}^{-15/32}{\rm\ K} .
\end{equation}
Shallow X-ray heating will dominate internal viscous heating in
setting the disk surface temperature when $T_x\gtrsim T_0$, which
occurs at radii exceeding roughly $10^9$ cm.  

If $T_x^4 \gtrsim \tau T_0^4$, then X-ray heating will also dominate
internal heating in setting the {\em central} temperature of the disk
(Lyutyi \& Sunyaev 1976; Spruit 1995).  Here, $\tau$ is the disk's
optical depth, given by $\tau= 34\ \alpha^{-4/5}\, \mu^{1/4}\, \dot
M_{-10}^{1/5}$ for a standard $\alpha$-disk (Frank et al. 1992).  In
this region of deep X-ray heating, the vertical structure of the
disk will not obey the $H\propto r^{9/8}$ relationship in equation
(\ref{eq-h98}), but will instead follow a $H\propto r^{9/7}$ power law
(e.g., Cunningham 1976; Vrtilek et al. 1990; Arons \& King 1993).
However, deep X-ray heating will dominate over shallow heating only
for radii exceeding roughly $2\times 10^{10}$ cm.  In a binary as
compact as 4U 1626--67, the disk will terminate before deep heating
becomes dominant, so that a combination of internal viscous heating
and shallow X-ray heating [equation (12) along with equations (11) and
(15)] should be valid over the entire disk.

Accounting for the presence of an X-ray heated accretion disk, it is
clear that an X-ray heated low-mass white dwarf at a distance of 1 kpc
is untenable for 4U 1626--67, since the expected optical emission from
the disk would be considerably brighter than the measured
fluxes. Instead, it is more appropriate to consider the opposite case
of the disk providing most of the optical luminosity by fitting
equations (10) and (12) along with equations (11) and (15) to the
measured optical photometry.   The free parameters for the fit are
$D$ (or, equivalently, $L_x$) and $\eta_d$.  As is evident from Figure
5,  the X-ray heated disk model can easily fit the data. 

However, as shown in Figure 6, the optical data do not provide a
strong joint constraint on $D$ and $\eta_d$. 
Ultraviolet observations will eventually provide a much stronger
constraint, since the UV spectrum from an X-ray heated disk is much
more sensitive to the values of $L_x$ and $\eta_d$.  In the absence of
ultraviolet observations, it should be noted that de Jong, van
Paradijs, \& Augusteijn (1996) have concluded that LMXB accretion
disks have a very high effective X-ray albedo ($\eta_d\gtrsim 
0.9$), based on observations of 4U~1254--69, 4U~1755--33, and
Sco~X-1.  If this is the case, then the optical data
constrain the distance to $5 \lesssim D \lesssim 13$ kpc (95\%
confidence), corresponding to an X-ray luminosity of $7\times 10^{36}
\lesssim L_x \lesssim 5\times 10^{37}$ erg s$^{-1}$.

\acknowledgements{I am grateful to John Grunsfeld for first
interesting me in this project. It is a pleasure to thank Andy Layden,
Ramon Galvez, Patricio Ugarte, and the CTIO staff for their
assistance and excellent support. I gratefully acknowledge a generous
thesis observing travel award from CTIO. I thank Lars Bildsten, Josh
Grindlay, Arlo Landolt, Al Levine, Rob Nelson, Tom Prince, and Brian
Vaughan for useful discussions. This work was funded in part by a NASA
GSRP Graduate Fellowship under grant NGT-51184, and by a NASA Compton
Postdoctoral Fellowship under grant NAG 5-3109.}


\pagebreak

{\small
\begin{deluxetable}{lcccccl}
\tablewidth{6.8in}
\tablecaption{Long Observations of 4U 1626--67}
\tablehead{
 &  & \colhead{Start time}  & \colhead{Duration} &   & 
     \colhead{High-frequency} & \\
\colhead{Run ID} & \colhead{Date} & \colhead{(UT)}& \colhead{(hr)} & 
     \colhead{Filter} & \colhead{contaminant?\tablenotemark{a}} & 
     \colhead{Pulsations?}}
\startdata
15U & 1995 May 26  & 0408 & 2.0 & $U$ &  No & Yes (strong detection) \\
17B & 1995 May 26  & 0638 & 1.8 & $B$ &  Yes& Yes (strong detection) \\
21I & 1995 May 27  & 0018 & 1.7 & $I$ &  No & No (clouds) \\
23R & 1995 May 27  & 0308 & 1.9 & $R$ &  No & Yes (strong detection) \\
25V2& 1995 May 27  & 0604 & 0.6 & $V$ &  Yes& No (clouds) \\
29U & 1995 May 28  & 0001 & 0.7 & $U$ &  Yes& No (clouds) \\
30U & 1995 May 28  & 0046 & 1.0 & $U$ &  Yes& Yes \\
32V & 1995 May 28  & 0224 & 1.0 & $V$ &  Yes& Yes \\
35U & 1995 May 28  & 0709 & 1.0 & $U$ &  Yes& Yes \\
36I & 1995 May 28  & 0813 & 1.0 & $I$ &  Yes& Yes (weak detection) \\
\enddata
\tablenotetext{a}{Instrumental signals at harmonics of 1.75 Hz and 60 Hz}
\end{deluxetable}
}

\begin{deluxetable}{lccccc}
\tablecaption{Optical Photometry of 4U 1626--67}
\tablehead{
\colhead{Quantity} & \colhead{$U$} &\colhead{$B$} &\colhead{$V$} &
  \colhead{$R$} &\colhead{$I$}}
\startdata
Source count rate (count s$^{-1}$) &  
 137(2) & 122(2) & 111(4) &177(3) & \nodata \nl
Pulsed  count rate (count s$^{-1}$) &
 7.9(3) & 6.9(3) & 7.0(5) & 11.8(5) & 25(2) \nl
Sky count rate (count s$^{-1}$) &
 168(2) & 182(2) & 317(4) & 785(3) & 5405(30) \nl
Magnitude & 
 17.50(29) & 18.70(24) & 18.68(15) &18.68(26) & \nodata \nl
Pulsed fraction & 
  5.8(2)\% & 5.7(3)\% & 6.3(5)\% & 6.7(3)\% & \nodata \nl
QPO fractional RMS strength & 3.4\%  & 2.7\%  & \nodata & 1.9\% &\nodata \nl
QPO centroid frequency (Hz) & 0.048 & 0.049  & \nodata & 0.049 &\nodata \nl
QPO FWHM (Hz) & 0.013  & 0.012  & \nodata & 0.009 &\nodata \nl
\enddata
\end{deluxetable}

\clearpage
\centerline{FIGURE CAPTIONS}

\noindent FIGURE 1: Optical pulse profiles as a function of wavelength
for 4U 1626--67. Also shown is the 20--60 keV X-ray pulse profile
measured contemporaneously by BATSE. Two pulses are shown in each
panel, and the pulse phase is plotted relative to the intensity
minimum of the X-ray pulse.  For the $I$-band optical and BATSE X-ray
measurements, the intensity of the unpulsed (DC) component of the
source emission was not measured. 

\noindent FIGURE 2: Fourier power density spectrum of a 2-hour $U$-band
observation (run 15U) of 4U 1626--67, normalized relative to the mean
source power.  The dotted line indicates the noise power level due to
Poisson counting statistics, which has been subtracted off. A 0.048 Hz
quasi-periodic oscillation with 4.9\% fractional RMS amplitude is
clearly detected, as is the 7.66-s coherent pulsation due to the
neutron star's rotation. An underlying $1/\nu$ spectrum is measured on
time scales $\lesssim 10$ s. 

\noindent FIGURE 3: Fourier power density spectrum of a 71 ks 2--6 keV {\em
RXTE}/PCA observation of 4U 1626--67 on 1996 Feb 11, normalized
relative to the mean X-ray source power.  The dotted line indicates
the noise power level due to Poisson counting statistics, which has
been subtracted off.   Although this power spectrum cannot be directly
compared to the optical power spectrum in Figure 2 (since they were
not acquired contemporaneously), the overall similarities are obvious
and bolster the idea that the optical emission is dominated by rapid X-ray
reprocessing.  Several higher harmonics of the 0.1304 Hz coherent
pulsation are observed due to the highly non-sinusoidal X-ray pulse
shape in the 2--6 keV band. 

\noindent FIGURE 4: A 5$\times$ over-resolved Fourier power spectrum
of 4U 1626--67 in the vicinity of the coherent 7.66-s pulsation,
normalized relative to the local noise power. The solid line is 
the average of three separate power spectra of 1.8-hr
observations. The dotted line is a model of the main pulsation's
expected sidelobe due to the finite length of the power spectrum. This
model is able to explain the symmetric sidelobes on either side of the
main pulsation.  A residual sideband downshifted 0.395(15) mHz from
the main pulsation (25\% relative amplitude) is also present, with a
statistical significance of 4$\sigma$.  This frequency shift is
consistent with a $2530\pm 100$ s prograde binary orbit. 

\noindent FIGURE 5:  Optical ({\em UBVRI}) photometry of 4U 1626--67
on 1995 May 26.  The solid curve indicates a typical X-ray heated disk
solution.  The particular solution shown assumed $\eta_d=0.95$,
$D=8.5$ kpc, $r_{\rm in}=6.5\times 10^8$ cm, and $r_{\rm out}=2\times
10^{10}$ cm.  However, a range of values for $\eta_d$ and $D$ provide
a good fit to the data (see Figure 6).

\noindent FIGURE 6: Distance to 4U 1626--67 as a function of effective
X-ray albedo of the disk, assuming that all the optical flux is due to
an X-ray heated accretion disk with $r_{\rm in}=6.5\times 10^8$ cm,
and $r_{\rm out}=2\times 10^{10}$ cm.  The shaded region is consistent
with the optical photometric data at the 95\%-confidence level.  Based
on observations of other LMXBs, de Jong et al. (1996) conclude that
$\eta_d\gtrsim 0.9$.  If this is correct, then the solutions allowed
by the data must lie in the shaded region to the right of the vertical
dotted line. 

\pagebreak
\pagestyle{empty}
\thispagestyle{empty}
\begin{figure}
\centerline{Figure 1}
\centerline{\psfig{file=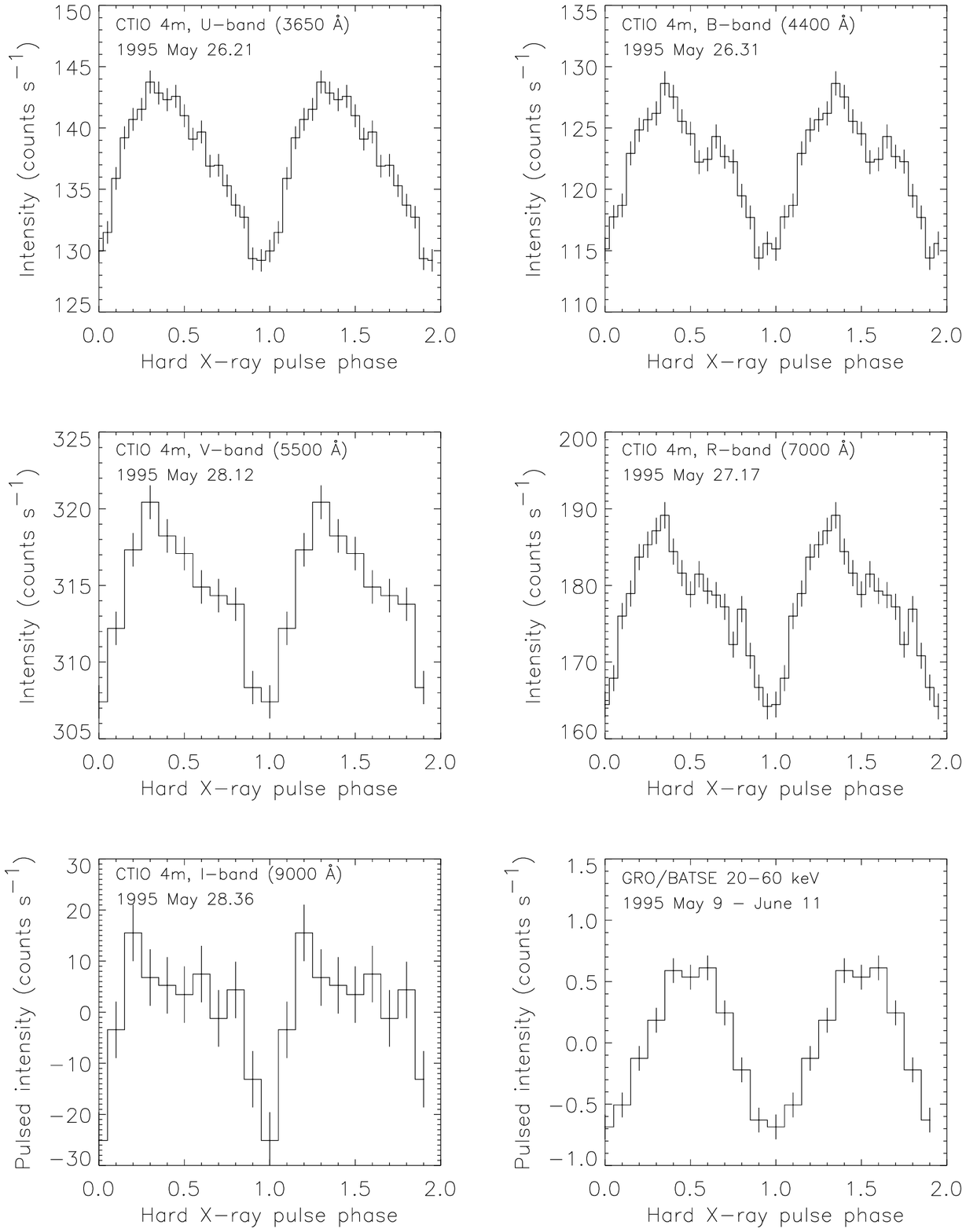}}
\end{figure}

\pagebreak
\pagestyle{empty}
\thispagestyle{empty}
\begin{figure}
\centerline{Figure 2}
\centerline{\psfig{file=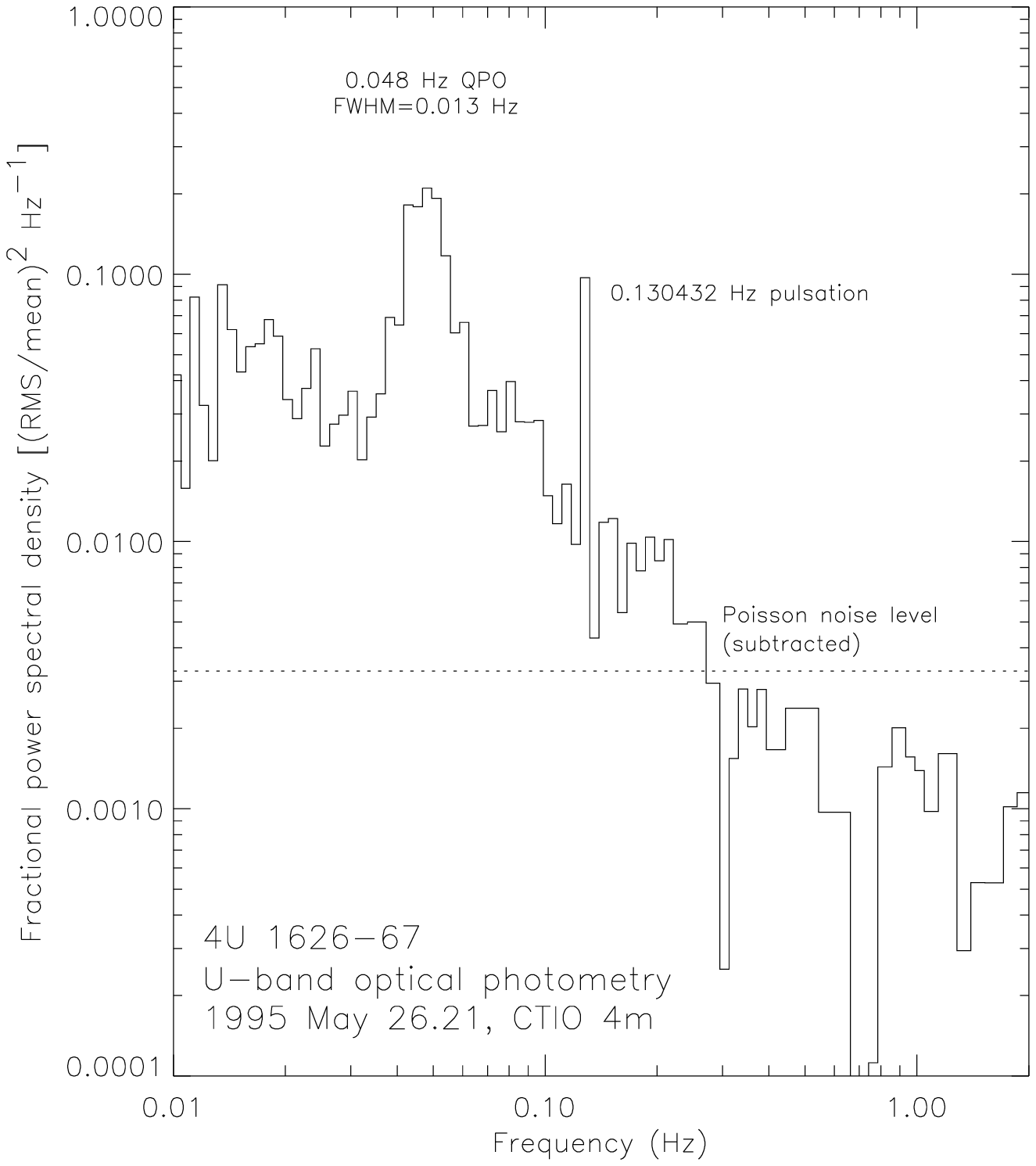}}
\end{figure}

\pagebreak
\pagestyle{empty}
\thispagestyle{empty}
\begin{figure}
\centerline{Figure 3}
\centerline{\psfig{file=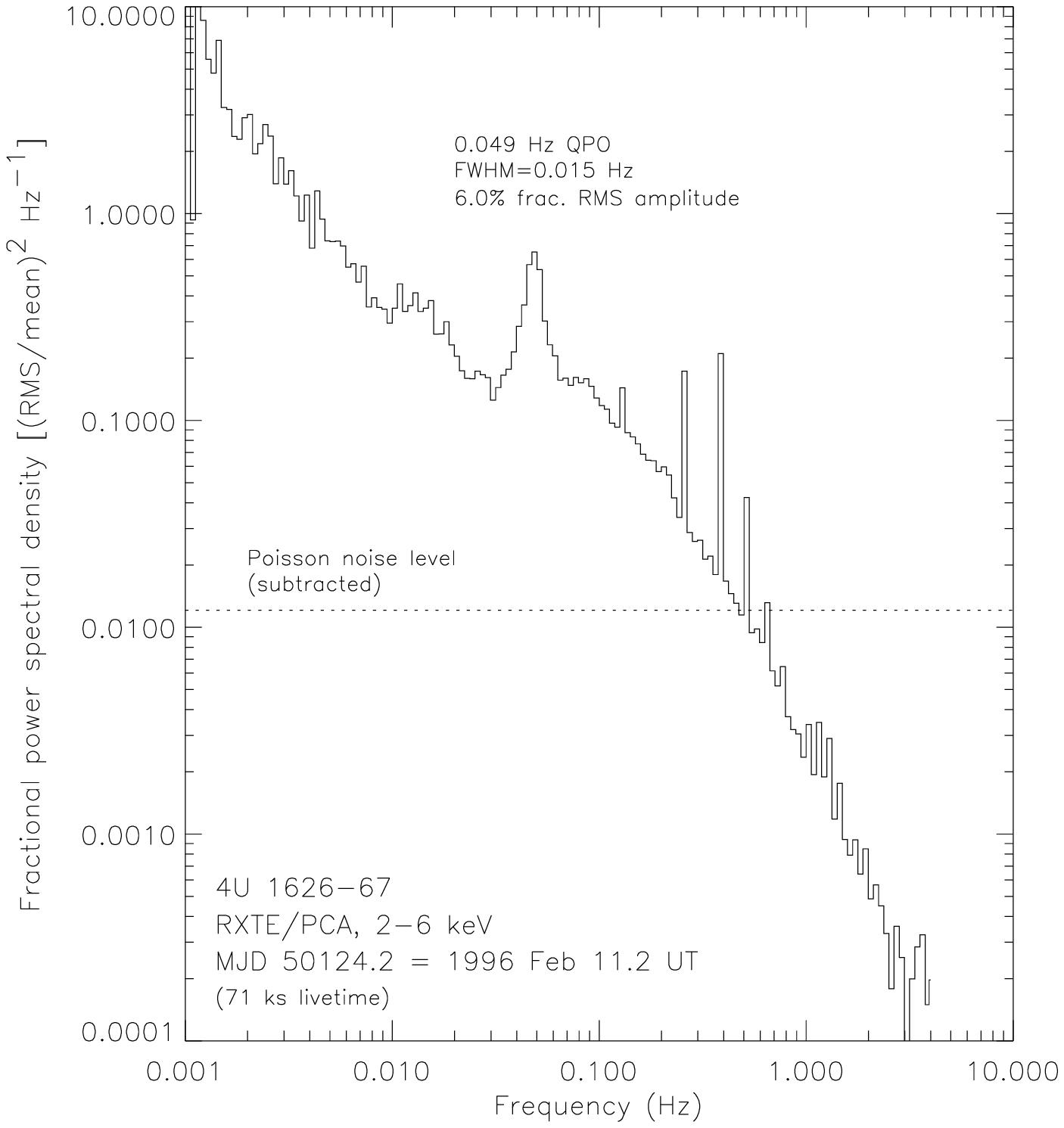}}
\end{figure}

\pagebreak
\pagestyle{empty}
\thispagestyle{empty}
\begin{figure}
\centerline{Figure 4}
\centerline{\psfig{file=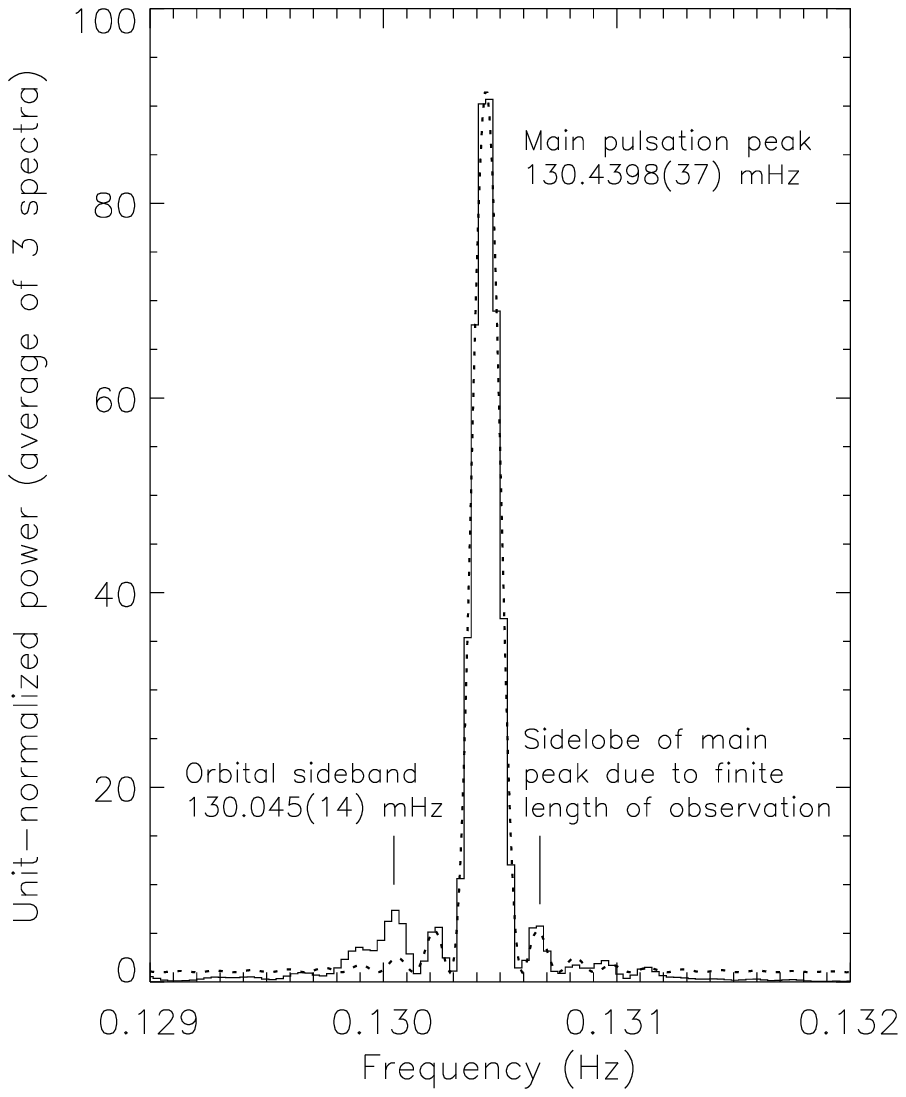}}
\end{figure}

\pagebreak
\pagestyle{empty}
\thispagestyle{empty}
\begin{figure}
\centerline{Figure 5}
\centerline{\psfig{file=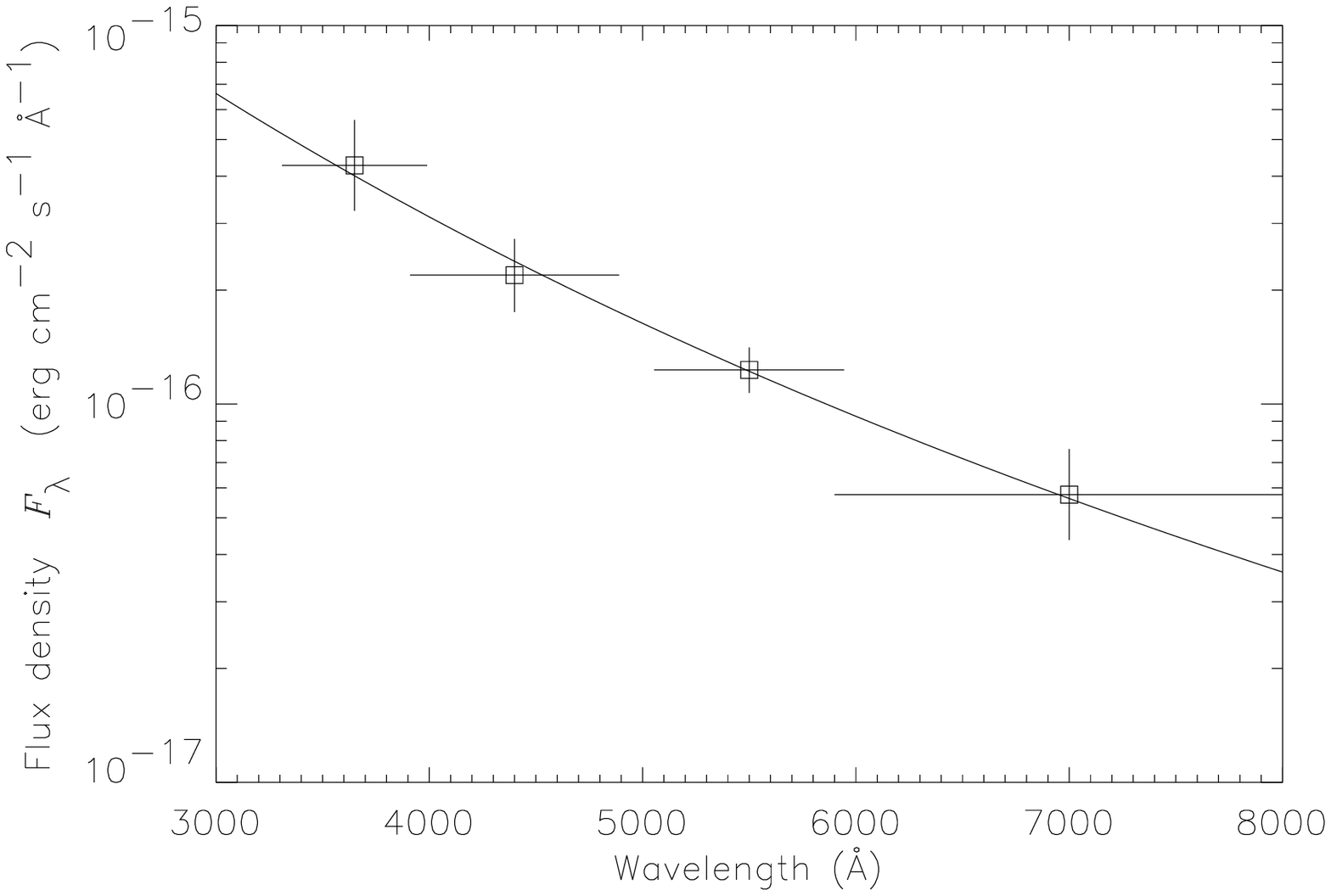}}
\end{figure}

\pagebreak
\pagestyle{empty}
\thispagestyle{empty}
\begin{figure}
\centerline{Figure 6}
\centerline{\psfig{file=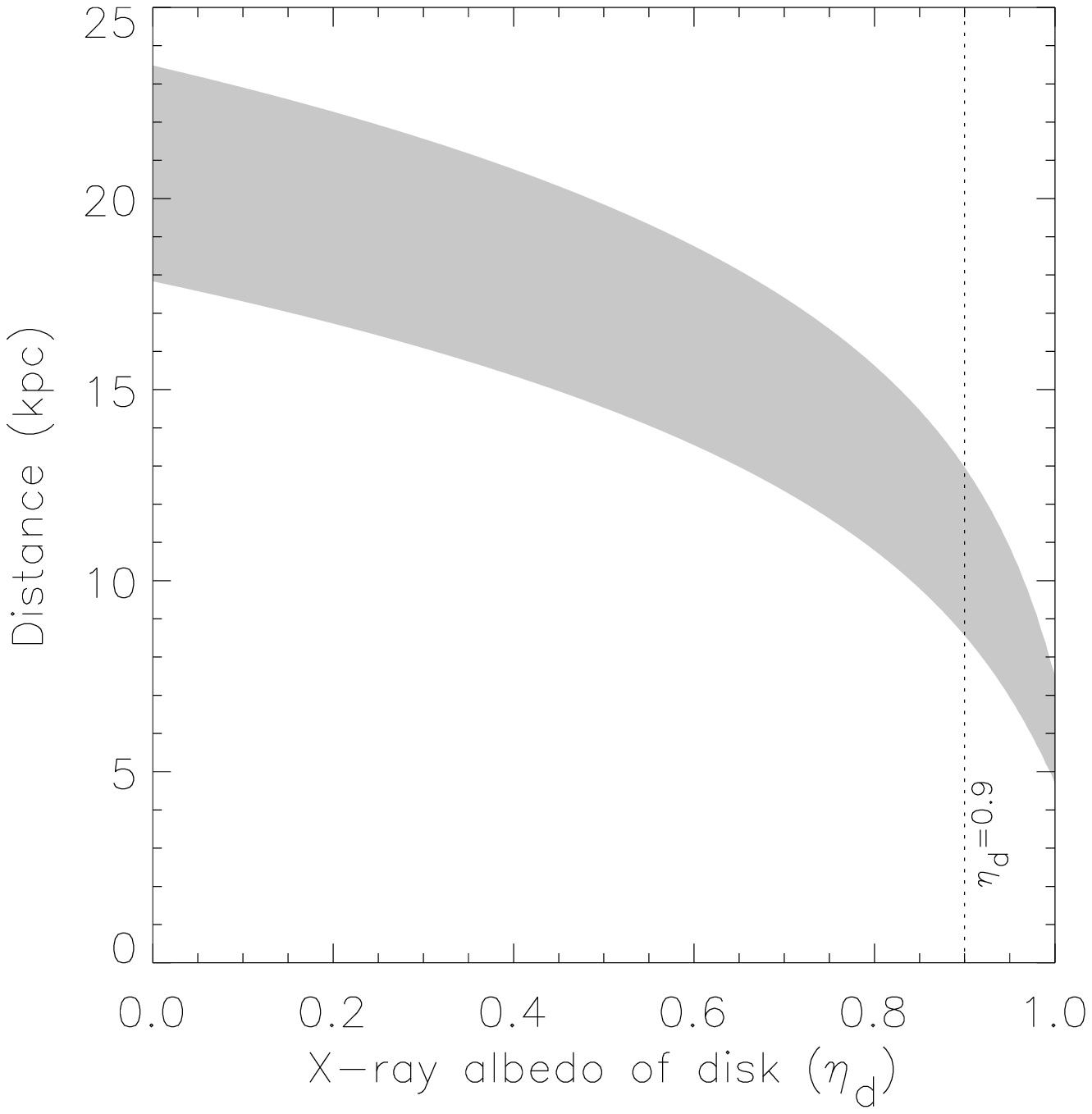}}
\end{figure}


\begin{thebibliography}{}
\parskip=0pt
\parsep=0pt
\itemsep=0pt


\bibitem[Abbott 1982]{Abbott82}
Abbott, D.~C. 1982, ApJ, 259, 282

\bibitem[Alpar \& Shaham 1985]{Alpar85}
Alpar, M.~A. \& Shaham, J. 1985, Nature, 316, 239

\bibitem[Anderson et al.  1997]{Anderson97}
Anderson, S.~F., Margon, B., Deutsch, E.~W., Downes, R.~A., \& Allen,
R.~G. 1997, ApJ, 482, L69

\bibitem[Angelini et al.  1995]{Angelini95}
Angelini, L., White, N.~E., Nagase, F., Kallman, T.~R., Yoshida, A.,
Takeshima, T., Becker, C.~M., \& Paerels, F. 1995, ApJ, 449, L41

\bibitem[Arons \& King  1996]{Arons96}
Arons, J. \& King, I.~R. 1996, ApJ, 413, L121

\bibitem[Beehler \& Lombardi 1990]{Beehler90}
Beehler, R.~E. \& Lombardi, M.~A. 1990, NIST Time and Frequency
Services, National Institute of Standards and Technology Publication
SP-432 

\bibitem[Burstein \& Heiles 1982]{Burstein82}
Burstein, D. \& Heiles, C. 1982, AJ, 87, 1165

\bibitem[Chakrabarty 1996]{Chak96}
Chakrabarty, D. 1996, Ph.D. thesis, California Institute of Technology

\bibitem[Chakrabarty et al. 1997]{Chak97}
Chakrabarty, D. et al. 1997, ApJ, 474, 414

\bibitem[Chester 1979]{Chester79}
Chester, T.~J. 1979, ApJ, 227, 569

\bibitem[Cunningham 1976]{Cunningham76}
Cunningham, C. 1976, ApJ, 208, 534

\bibitem[Davidsen et al. 1972]{Davidsen72}
Davidsen, A., Henry, J.~P., Middleditch, J., \& Smith, H.~E. 1972,
ApJ, 177, L97

\bibitem[de Jong, van Paradijs, \& Augusteijn 1996]{deJong96}
de Jong, J.~A., van Paradijs, J., \& Augusteijn, T. 1996, A\&A, 314, 484

\bibitem[Eggleton 1983]{Eggleton83}
Eggleton, P. P. 1983, ApJ, 268, 368

\bibitem[Frank et al. 1992]{Frank92}
Frank, J., King, A., \& Raine, D. 1992, Accretion Power in
Astrophysics, 2nd ed. (Cambridge: Cambridge U. Press)


\bibitem[Graham 1982]{Graham82}
Graham, J.~A. 1982, PASP, 94, 244

\bibitem[Grindlay 1978]{Grindlay78}
Grindlay, J.~E. 1978, ApJ, 225, 1001

\bibitem[Homer et al. 1996]{Homer96}
Homer, L., Charles, P.~A., Naylor, T., van Paradijs, J., Auri\`ere,
M., \& Koch-Miramond, L. 1996, MNRAS, 282, L37

\bibitem[Ilovaisky, Motch, \& Chevalier 1978]{Ilovaisky78}
Ilovaisky, S.~A., Motch, C., \& Chevalier, C. 1978, A\&A, 70, L19

\bibitem[Imamura et al. 1990]{Imamura90}
Imamura, J.~N., Kristian, J., Middleditch, J., \& Steiman-Cameron,
T.~Y. 1990, ApJ, 365, 312

\bibitem[Jablonski et al. 1997]{Jablonski97}
Jablonski, F.~J., Pereira, M.~G., Braga, J., \& Gneiding, C.~D. 1997,
ApJ, 482, L171

\bibitem[Jenkins \& Watts 1968]{Jenkins68}
Jenkins, G.~M. \& Watts, D.~G. 1968, Spectral Analysis and Its
Applications (San Francisco: Holden-Day)

\bibitem[Joss, Avni, \& Rappaport 1978]{Joss78}
Joss, P.~C., Avni, Y., \& Rappaport, S. 1978, ApJ, 221, 645

\bibitem[Kippenhahn \& Weigert 1990]{Kippenhahn90}
Kippenhahn, R. \& Weigert, A. 1990, Stellar Structure and Evolution
(Berlin: Springer-Verlag)

\bibitem[Lamb et al. 1985]{Lamb85}
Lamb, F.~K., Shibazaki, N., Alpar, M.~A., \& Shaham, J. 1985, Nature,
317, 681

\bibitem[Lewin, van Paradijs, \& Taam 1993]{Lewin93}
Lewin, W.~H.~G., van Paradijs, J., \& Taam, R.~E. 1993, Space
Sci. Rev., 62, 223


\bibitem[Levine et al.  1988]{Levine88}
Levine, A., Ma, C.~P., McClintock, J., Rappaport, S., van~der Klis, M., \&
Verbunt, F. 1988, ApJ, 327, 732

\bibitem[Li et al.  1980]{Li80}
Li, F.~K., Joss, P.~C., McClintock, J.~E., Rappaport, S., \& Wright, E.~L.
1980, ApJ, 240, 628

\bibitem[Lyutyi \& Sunyaev  1976]{Lyutyi76}
Lyutyi, V.~M. \& Sunyaev, R.~A. 1976, Sov. Astron., 20, 290

\bibitem[McClintock et al.  1977]{McClintock77}
McClintock, J.~E., Canizares, C.~R., Bradt, H.~V., Doxsey, R.~E., Jernigan,
J.~G., \& Hiltner, W.~A. 1977, Nature, 270, 320

\bibitem[McClintock et al.  1980]{McClintock80}
McClintock, J.~E., Canizares, C.~R., Li, F.~K., \& Grindlay,
J.~E. 1980, ApJ, 235, L81

\bibitem[Middleditch \& Nelson 1976]{Middleditch76}
Middleditch, J. \& Nelson, J. 1976, ApJ, 208, 567

\bibitem[Middleditch et al.  1981]{Middleditch81}
Middleditch, J., Mason, K.~O., Nelson, J.~E., \& White, N.~E. 1981,
ApJ, 244, 1001

\bibitem[Miyamoto et al. 1994]{Miyamoto94}
Miyamoto, S., Kitamoto, S., Iga, S., Hayashida, K., \& Terada, K.
1994, ApJ, 435, 398

\bibitem[Motch et al. 1983]{Motch83}
Motch, C., Ricketts, M.~J., Page, C.~G., Ilovaisky, S.~A., \&
Chevalier, C. 1983, A\&A, 119, 171

\bibitem[Motch, Ilovaisky, \& Chevalier 1982]{Motch82}
Motch, C., Ilovaisky, S.~A., \& Chevalier, C. 1982, A\&A, 109, L1

\bibitem[Nelson et al.  1986]{Nelson86}
Nelson, L.~A., Rappaport, S.~A., and Joss, P.~C. 1986, ApJ, 304, 231

\bibitem[Nelson et al. 1997]{Nelson97}
Nelson, R.~W. et al. 1997, in Accretion Phenomena and Associated
Outflows, ed. D. Wickramasinghe, L. Ferrario, \& G. Bicknell (San
Francisco: Astron. Soc. of the Pacific), in press 

\bibitem[Paczynski \& Sienkiewicz  1981]{Paczynski81}
Paczynski, B. \& Sienkiewicz 1981, ApJ, 248, L27

\bibitem[Pravdo et al. 1979]{Pravdo79}
Pravdo, S.~H. et al. 1979, ApJ, 231, 912

\bibitem[Press et al. 1992]{Press92}
Press, W.~H., Teukolsky, S.~A., Vetterling, W.~T., \& Flannery,
B.~P. 1992, Numerical Recipes in C: The Art of Scientific
Computing (2d ed.; Cambridge: Cambridge U. Press)

\bibitem[Ravenhall \& Pethick 1994]{Ravenhall94}
Ravenhall, D.~G. \& Pethick, C.~J. 1994, ApJ, 424, 846

\bibitem[Reynolds et al. 1997]{Reynolds97}
Reynolds, A.~P., Quaintrell, H., Still, M.~D., Roche, P.~D.,
Chakrabarty, D., \& Levine, S.~E. 1997, MNRAS, accepted for publication

\bibitem[Savonije et al. 1986]{Savonije86}
Savonije, G.~J., de Kool, M., \& van den Heuvel, E.~P.~J. 1986, A\&A,
155, 51

\bibitem[Shakura \& Sunyaev 1973]{Shakura73}
Shakura, N.~I. \& Sunyaev, R.~A. 1973, A\&A, 24, 337

\bibitem[Shapiro \& Teukolsky 1983]{Shapiro83}
Shapiro, S.~L. \& Teukolsky, S.~A 1983, Black Holes, White Dwarfs, and
Neutron Stars (New York: Wiley)

\bibitem[Shibazaki \& Lamb 1987]{Shibazaki87}
Shibazaki, N. \& Lamb, F.~K. 1987, ApJ, 318, 767

\bibitem[Shinoda et al.  1990]{Shinoda90}
Shinoda, K., Kii, T., Mitsuda, K., Nagase, F., Tanaka, Y., Makishima, K., \&
Shibazaki, N. 1990, PASJ, 42, L27

\bibitem[Spruit 1995]{Spruit95}
Spruit, H.~C. 1995, in The Lives of the Neutron Stars,
ed. M.~A. Alpar, U. Kiziloglu, \& J. van Paradijs (Dordrecht: Kluwer)

\bibitem[Standish et al. 1992]{Standish92}
Standish, E.~M., Newhall, X.~X., Williams, J.~G., \& Yeomans, D.~K.
1992, in Explanatory Supplement to the Astronomical Almanac, ed. P.~K.
Seidelmann (Mill Valley: University Science), 279

\bibitem[Tavani \& London 1993]{Tavani93}
Tavani, M. \& London, R. 1993, ApJ, 410, 281

\bibitem[van der Klis 1995]{VDK95}
van der Klis, M. 1995, in X-Ray Binaries, ed. W.~H.~G. Lewin, J. van
Paradijs, \& E.~P.~J. van den Heuvel (Cambridge: Cambridge
Univ. Press), 252 

\bibitem[van der Klis et al. 1987]{VDK87}
van der Klis, M., Jansen, F., van Paradijs, J., Lewin, W.~H.~G.,
Sztajno, M., \& Tr\"umper, J. 1987, ApJ, 313, L19

\bibitem[van Paradijs, \& McClintock 1995]{van95}
van Paradijs, J. \& McClintock, J.~E. 1995, in X-Ray
Binaries, ed. W.~H.~G. Lewin, J. van Paradijs, \& E.~P.~J. van den
Heuvel (Cambridge: Cambridge Univ. Press), 58

\bibitem[Verbunt \& van den Heuvel 1995]{Verbunt95}
Verbunt, F. \& van den Heuvel, E.~P.~J. 1995, in X-Ray
Binaries, ed. W.~H.~G. Lewin, J. van Paradijs, \& E.~P.~J. van den
Heuvel (Cambridge: Cambridge Univ. Press), 457

\bibitem[Verbunt, Wijers, \& Burm 1990]{Verbunt90}
Verbunt, F., Wijers, R. A. M.~J., \& Burm, H. M.~G. 1990, A\&A, 
234, 195

\bibitem[Vrtilek et al. 1990]{Vrtilek90}
Vrtilek, S.~D., Raymond, J.~C., Garcia, M.~R., Verbunt, F., \&
Hasinger, G. 1990, A\&A, 235, 162

\bibitem[Warner 1995]{Warner95}
Warner, B. 1995, Ap\&SS, 225, 249

\end{thebibliography}
\end{document}